\documentclass[12pt]{article}

\usepackage{graphicx}
\usepackage{amssymb}
\usepackage{amsmath}

\usepackage{slashed}

\newcommand{\TeV}{~\mathrm{TeV}}

\newcommand{\uonep}{\ensuremath{U(1)^\prime}}
\newcommand{\zp}{\ensuremath{Z^\prime}}
\newcommand{\gz}{\ensuremath{g_{z^\prime}}}
\newcommand{\SD}{S_{\scriptscriptstyle D}}

\begin{document}

\begin{titlepage}

\begin{flushright}
November 7, 2008
\end{flushright}

\vspace{0.7cm}
\begin{center}
\Large\bf\boldmath
Scalar Potentials and Accidental Symmetries in Supersymmetric $\uonep$ Models
\unboldmath
\end{center}

\vspace{0.8cm}
\begin{center}
{\sc Paul Langacker$^{(a)}$, Gil Paz$^{(a)}$, and Itay Yavin$^{(b)}$} \\
\vspace{0.7cm}
{\sl ${}^a$\, School of Natural Sciences, Institute for Advanced Study\\  Princeton, NJ 08540, U.S.A\\[0.3cm]
${}^b$\, Department of Physics, Princeton University \\Princeton, NJ 08544, U.S.A }
\end{center}

\vspace{1.0cm}
\begin{abstract}
\vspace{0.2cm}
\noindent
We address two closely related
problems associated with the singlet scalars' potential that are
often present in supersymmetric $\uonep$ models, especially those
which maintain the gauge unification of the MSSM in a simple way.
The first is the possibility  of an accidental global symmetry which
results in a light Goldstone boson. The second is the  problem of
generating a vacuum expectation value for more than one field
without reintroducing the $\mu$ problem. We give sufficient
conditions for addressing both issues and provide a concrete example
to generate them.
\end{abstract}
\vfil

\end{titlepage}

\section{Introduction}

Extensions of the standard model (SM) and the Minimal Supersymmetric
SM (MSSM) frequently involve additional abelian gauge symmetries,
often at the TeV scale (for reviews,
see~\cite{Leike:1998wr,Rizzo:2006nw,Langacker:2008yv}). The primary
motivation for considering such scenarios is top-down, i.e., because
so many extensions involve larger gauge symmetries which often leave
an abelian remnant when they are broken. Another motivation is that
many supersymmetric \uonep\ models provide an elegant solution of the
$\mu$ problem~\cite{Kim:1983dt}, by forbidding an elementary $\mu$
term but allowing a dynamical $\mu$ to be generated by the vacuum
expectation value (VEV) of a SM-singlet field charged under the
$\uonep$~\cite{Suematsu:1994qm,Cvetic:1995rj,Cvetic:1997ky} (other
models with a dynamical $\mu$ are reviewed
in~\cite{Accomando:2006ga}). However, like most extensions of the SM
or MSSM, the seemingly innocent addition of an abelian group brings
with it difficulties and complications in constructing
phenomenologically viable models.

Aside from hypercharge, $B-L$ (which does {\em not} forbid an
elementary $\mu$) is the only abelian family-universal global symmetry
of the Standard Model (defined to include three right-handed
neutrinos) that is anomaly free. Thus, in order to simultaneously
implement the $\uonep$ solution to the $\mu$ problem and satisfy all
the anomaly cancellation conditions (ACC), one is forced to introduce
new chiral exotic matter when extending the gauge structure to include
an additional abelian group\footnote{One may view this as nuisance and
  a good reason to dismiss extra abelian groups other than $B-L$ at
  low energies. Alternatively, one can view it as a boon since the
  anomaly structure \textit{predicts} new exotic matter which may be
  observed alongside the $\zp$. Others implications of $\uonep$ models
  are reviewed in~\cite{Langacker:2008yv}.}.  It is usually assumed to
be quasi-chiral, i.e., vector-like with respect to the SM gauge group
but chiral under the \uonep. Being charged under the SM gauge groups
forbids this exotic matter from being
light. 

Unless they come in complete $SU(5)$-type multiplets, the exotic
matter typically ruins the simple form of unification found in the
MSSM\footnote{The model considered in Ref.
  \cite{Langacker:2007ac,Langacker:2008ip} is an example where all the
  ACC are satisfied, but the field content spoils unification.},
although unification can often be restored by adding ad hoc adjoint or
vector-like fields at the TeV or intermediate scales. For example,
$E_6$-type models~\cite{Hewett:1988xc} can accommodate all the needed
exotics and Higgs fields in three 27-plets.  However, gauge
unification is not respected unless one adds an ad hoc vector pair of
$SU(2)$ doublets, e.g., from an incomplete $27+27^\ast$, and
generating their masses introduces a new vector doublet version of the
$\mu$ problem~\cite{Langacker:1998tc}.  Another class of
models~\cite{Erler:2000wu,Morrissey:2005uz} {\em is} consistent with
simple gauge unification. This is achieved by starting with the MSSM
fields and adding to it additional multiplets that transform like
complete $SU(5)$ multiplets under the SM gauge group. However, not all
of the fields have the same $\uonep$ charges, i.e., the assignments do
not descend from an underlying $SU(5)\times \uonep$.

The last ingredient needed are SM singlet fields which are charged
under the new abelian gauge-group. These singlets are required to
break the $\uonep$ symmetry, give the exotics large enough masses, and
satisfy all the linear and cubic ACC for the $\uonep$ group. Thus, the
singlet sector is given several duties and the field content is
constrained.

In supersymmetric theories these provisions are further complicated by
the holomorphy of the superpotential and the special structure of the
resulting scalar potential. Together with charge conservation, the ACC
often fix the scalars' charges almost entirely. This rigidity,
exacerbated by holomorphy, allows few, if any, terms in the
superpotential. The resulting scalar potential then suffers from two
generic problems. First, accidental symmetries are present which once
broken lead to an axion-like boson in the spectrum which is
experimentally excluded. Second, the true vacuum is often such that
not all the scalars develop a VEV.  This is not a difficulty for those
scalars which are only needed for the ACC, but is a phenomenological
disaster\footnote{Frequently, the fermionic components of some of the
  singlets remain massless or very light, even when there are no
  issues of accidental symmetries or when the scalar does not acquire
  a VEV. Such fermions are similar to sterile neutrinos, and do not
  cause any major problems as long as they do not mix with ordinary
  neutrinos and some non-trivial but not overly stringent conditions
  are met to avoid astrophysical and cosmological
  difficulties~\cite{Langacker:2008yv}. Light singlet fermions can
  even be helpful in allowing sufficiently rapid decays of exotic
  particles via higher-dimensional operators~\cite{Kang:2007ib}.}  for
those which are needed to generate masses for the exotics.

Both of these problems are often easier to resolve once we allow for
bilinear terms. However, we would like to avoid reintroducing the
usual $\mu$-problem in the singlet sector. There are actually two
aspects to the $\mu$ problem. The first is to {\em prevent} the
presence of unacceptably large bilinear terms, e.g., at the string or
GUT scale. This is not difficult, as one can always imagine that
constraints from some underlying string theory, for example, force the
bilinear to vanish initially. More difficult is the problem of
introducing the bilinear and its associated soft term at the {\em
  right} scale. What the ``right scale'' is depends on the problem we
are trying to solve. As far the accidental symmetries are concerned,
the bilinear needs to be large enough to avoid an axion in the
spectrum. This is only a lower bound on the size of the bilinear and
does not represent a serious concern. The more serious problem arises
when attempting to utilize bilinears to spontaneously break a
symmetry. In that case, the relevant ``$\mu$'' and ``$B\mu$'' terms
have to be at the same scale as the other soft parameters. In this
paper, we refrain from using bilinears in order to achieve spontaneous
symmetry breaking. In that strict sense, we do not reintroduce the
``$\mu$-problem''.

In this letter we address these issues, the presence of accidental
symmetries and the generation of multiple VEVs.
We begin with a somewhat general discussion of the points at hand,
although it is difficult to produce any rigorous proofs for the most
general case. We also provide some specific examples where more
definitive statements can be made. Ultimately, our purpose is to draw
attention to some aspects of $\zp$ model building which are often
neglected.

\section{General Remarks}

\subsection{Accidental phase symmetries}
The problem of accidental symmetries arises whenever we have two or
more SM singlet superfields with different nonzero \uonep\ charges, as
are required in the models in~\cite{Erler:2000wu} to generate masses
for all of the exotics. To illustrate the problem, assume first that
the singlet sector has no superpotential and the scalar potential for
the $N$ singlet scalar components contains only soft mass and
$D$-terms,
\begin{equation}\label{eq:Vgeneral}
V(S_1,...,S_N)=\sum_i m_i^2\, |S_i|^2+\frac{\gz^2}2\left(\sum_i
Q_i\,|S_i|^2 \right)^2.
\end{equation}
Here $\gz$ is the $\uonep$ gauge coupling constant and $Q_i$ and
$m_i^2$ are the $\uonep$ charge and soft mass-squared of the field
$S_i$, respectively. This potential has $N-1$ ``accidental" global
$U(1)$ symmetries. When the scalar fields develop VEVs these
symmetries will be broken, resulting in $N-1$ massless Nambu-Goldstone
bosons (NGB). These accidental symmetries are generically anomalous
with respect to the SM gauge group. Therefore, one linear combination
of the NGBs is an axion~\cite{Georgi:1986df} as it receives a small
mass of order $\Lambda_{\rm QCD}^2/f$, where $f$ is a typical scalar
VEV, and the rest are massless. The presence of such massless scalars
is excluded by constraints on the existence of a fifth force (see, for
example, Refs. \cite{Long:2003ta,Adelberger:2003zx} for recent
reviews).  As for the axion state, for $f\sim 100\TeV$ the mass would
be around $100~\mathrm{eV}$. Such light scalars are excluded by axion
searches, which require that the axion mass should be below
$10~\mathrm{meV}$ \cite{Amsler:2008zz}.  One can attempt to break the
symmetry through a higher dimensional operator suppressed by some mass
scale, $\Lambda_{\,\rm cutoff}$.  The axion mass in this case would be
of the order $f^2/\Lambda_{\,\rm cutoff}$.  But even for
$\Lambda_{\,\rm cutoff}=M_p$, where $M_p$ is the Planck scale, the
axion mass would be too large.  If, on the other hand, we wish to
generate a large mass via this higher dimensional operator, e.g.
larger than $\sim$ MeV to avoid astrophysical and laboratory bounds,
we must take $\Lambda_{\,\rm cutoff}$ to be much smaller than the GUT
scale.

To avoid this situation, we need to have $N-1$ \textit{linearly
  independent} terms in the superpotential. Since the chiral
supermultiplets are charged under $\uonep$, no linear term can be
written.  Ideally, one would like to have only cubic terms, but we can
allow for bilinear terms as long as they do not reintroduce the $\mu$
problem. Thus, such terms will not be used to achieve any particular
vacuum structure, but they can be utilized to break accidental
symmetries. A bilinear term of order $F/M_p$, where $F$ is the
auxiliary field's VEV in the hidden sector,
can be generated by the Giudice-Masiero \cite{Giudice:1988yz}
mechanism.  This can easily give the NGB a large enough mass, e.g. of
order MeV, even for a relatively low $F$, such as is found in models
of gauge mediation (for a review, see e.g., \cite{Chung:2003fi}), thus
avoiding any current bounds on light scalars. In contrast, the
Giudice-Masiero mechanism is only useful for the original Higgs $\mu$
problem for relatively large $F$, such as is found in supergravity
mediation.

Regardless of the mechanism that mediates SUSY breaking, such
accidental symmetry breaking terms in the superpotential would
radiatively generate $A$ and $B$ terms, which together with the $F$
terms are enough to generate masses for all the would-be-axions.

\medskip
\textbf{Cubic and bilinear terms}
\medskip

The typical situation one encounters is that we have $k$ singlet
fields and we would like to add $l$ singlet fields, none of which has
zero $\uonep$ charge, such that we have $k+l-1$ linearly independent
terms in the superpotential.  The question one might ask is,
can we always find such $l$ fields, for large enough $l$? If we do,
can we use only cubic terms? The answer of course will depend on the
$\uonep$ charges of the given $k$ fields. But we can make some fairly
general statements, assuming that the charges are ``small" rational
numbers.

In general the charges of the singlet fields can be divided into
equivalence classes according to their congruence modulo 3. The terms
in the superpotential can also be written as equations for the
charges. The possible cubic terms $S_i^2S_j$ and $S_iS_jS_m$ can be
written as $2Q_{i}+Q_{j}=0$ and $Q_{i}+Q_{j}+Q_{m}=0$, respectively. A
possible bilinear term can be written as $Q_{i}+Q_{j}=0$.  Equations
of the form $2Q_{i}+Q_{j}=0$ can only ``connect" charges which are
congruent modulo 3. Equations of the type $Q_{i}+Q_{j}+Q_{m}=0$
connect either charges which are congruent modulo 3, or connect
charges from three \textit{different} equivalence classes, e.g.,
$Q_{i}$ is 0 mod 3, $Q_{j}$ is 1 mod 3, and $Q_{m}$ is 2 mod 3.
Equations of the form $Q_{i}+Q_{j}=0$ can have solutions only if
$Q_{i}$ is 1 mod 3 and $Q_{j}$ is 2 mod 3, or if both $Q_{i}$ and
$Q_{j}$ are 0 mod 3.

It is difficult to break the accidental symmetries with a small number
of fields using only the cubic terms if the $\uonep$ charges of the
initial set belong to different equivalence classes.  That is because
the terms $S_iS_jS_m$ have to connect three different equivalence
classes and there must be enough of them.  There is no such
restriction for bilinears.
We will see explicit examples for this phenomena in the next section.
On the other hand, if most (or all) of the charges belong to the same
equivalence class, we would not expect to have a substantial
difference between bilinear and cubic terms and one can probably use
only cubic terms.

\mbox{}

\subsection{Multiple scalar condensation}
As far as VEVs of the different fields are concerned, cubic terms
are often sufficient to ensure that all the relevant scalars
develop a VEV. While it is difficult to make any general statements
about the minimization of such potentials and the different possible
phases encountered, it is fairly easy to understand how one might
generate VEVs for multiple fields. Let us first assume that the
$\uonep$ gauge coupling $\gz$ is somewhat larger than any of the
cubic couplings, denoted generically by $y$. Then, neglecting the cubic
couplings, the scalar potential in (\ref{eq:Vgeneral}) minimizes
according to,
\begin{equation}\label{eqn:DtermCont}
|S_i|\left[m_i^2 + \gz^2\,Q_i\left(\sum_j Q_j |S_j|^2\right) \right]
= 0 ,
\end{equation}
where $i$ and $j$ run over all the scalars involved. At this point,
one of two things might happen. The first possibility is that one (and
only one) of the fields develops a VEV, provided its mass-squared was
driven negative by the renormalization group equations (RGEs). If
there are several fields with negative mass-squared, the one for which
$|m_i^2/Q_i|$ is the largest, will develop the VEV. This will remain
the minimum even after including the $F$ terms generated by the
superpotential cubic terms, as long as no flat direction is present.
Other fields may then develop VEVs through $A$-terms which add linear
terms to the scalar potential (after VEV insertions).

The second possibility is that a flat direction is present. If, for
example, any two of the fields $S_i$ and $S_j$ with opposite-sign
charges have $|Q_j| m_i^2 + |Q_i| m_j^2 < 0$ then we have a ``runaway"
direction, i.e., $V \to -\infty$, for $|Q_i| |S_i|^2 = |Q_j| |S_j|^2
\to\infty$ and $ |S_k|^2 =0,\,k\neq i,j $ . This is desirable, because
once we include the $F$-terms in the potential, these fields will be
stabilized at finite values\footnote{This mechanism is utilized in the
  secluded sector models~\cite{Erler:2002pr}, which allow $M_{\zp}$ to
  be considerably above the effective $\mu$. Even if the potential is
  not stabilized by renormalizable level $F$-terms, it is may be
  stabilized at intermediate scales because of the running $m^2$ or
  due to higher dimensional operators~\cite{Cleaver:1997nj}.}. The
presence of these flat directions lingers in the VEVs being
proportional to $1/y^2$. Then again, upon including the $A$-terms,
many other fields may develop a VEV as discussed above.
Clearly, one cannot count on too many $A$-terms without imposing
fairly stringent constraints on the different charge assignments.

If we relax the condition that $\gz$ is larger than all of the cubic
couplings, we can have two possibilities depending on the presence
of $F$ terms in the scalar potential of the form
$|y|^2\,|S_i|^2\,|S_j|^2$ with $y>\gz$. If such terms are
not present, the situation is identical to the $\gz>y$ case
considered above. If such terms are present, we can generate
positive effective mass terms for $S_j$ via VEV insertion, i.e.,
$m_j^2=|y|^2\,|S_i|^2$. Such a positive effective
mass term may overcome the negative soft mass term and restabilize
the point $|S_j|=0$. We will see examples for both of
these cases in the next section.

\section{An Example: Erler's Model}

As a concrete example, illustrating the challenges and their
resolutions discussed above, we consider one of the models presented
in Ref. \cite{Erler:2000wu}, which constructed supersymmetric $\uonep$
models which solve the $\mu$ problem in the MSSM while maintaining
simple gauge unification.

The charge assignments in the example we consider, given in Table
(\ref{tbl:gaugeunif}), are slightly
generalized~\cite{Langacker:2008yv} from the model
in~\cite{Erler:2000wu}. The free parameters $x$,$y$ and $z$ are
determined through the cubic anomaly cancellation condition. The
singlet fields $S$, $\SD$ and $S_{\scriptscriptstyle L}$, which are
responsible for generating the $\mu$ term and giving the exotics a
mass, are independent of $x$,$y$ and $z$. We will therefore ignore
these parameters since they carry no significance in our analysis.

\begin{table}[ht]
\begin{center}
\begin{tabular}{|c|c|| c| c|}
\hline &$\uonep$ charge& & $\uonep$ charge \\
\hline
$Q$ & $y$           & $H_u$ & $x$  \\
$u^c$ & $-x-y$       & $H_d$ & $-1-x$  \\
$d^c$ & $1+x-y$       & $\SD$ & $3/n_{\bf 55^\ast}$  \\
$L$ & $1-3y$           & $D_i$ & $z$  \\
$e^+$ & $x+3y$       & $D_i^c$ & $-3/n_{\bf 55^\ast}-z$  \\
$\nu^c$ & $-1-x+3y$    & $S_{\scriptscriptstyle L}$ & $2/n_{\bf 55^\ast}$  \\
$S$ & $1$         & $L_i$ &  $\frac{5-n_{\bf 55^\ast}}{4n_{\bf 55^\ast}}+x+3y+3z/2$  \\
&     & $L_i^c$ & $-2/n_{\bf 55^\ast}-Q_{L_i}$  \\
\hline
\end{tabular}
\end{center}
\caption{Examples of supersymmetric models consistent with minimal SM gauge
unification. $n_{\bf 55^\ast}$ is the number of pairs of  ${\bf 5} + {\bf 5^\ast}$.
$Q_S$ is taken to be 1. The free parameters are $Q_{H_u}\equiv x, Q_Q\equiv y,
Q_D\equiv z$.
\label{tbl:gaugeunif}}
\end{table}

\subsection{Singlets' scalar potential}
To generate an effective $\mu$ term and give masses to the exotics
$D_i$, $D_i^c$ and $L_i$, $L_i^c$, all three singlets, $S$, $\SD$ and
$S_{\scriptscriptstyle L}$ need to develop VEVs. Since they all carry
positive $\uonep$ charge, we cannot write any interaction terms in the
superpotential. Assuming SUSY is broken, the scalar potential will
involve only the soft mass terms and the $\uonep$ $D$-terms. In this
case, as discussed immediately after Eq. (\ref{eqn:DtermCont}), only
one of the fields will develop a VEV\footnote{We are ignoring the
  Higgs scalars contribution to the potential. This is certainly
  justified when the $\uonep$ is broken at a scale much higher than
  the EW scale.  Moreover, since the Higgs fields only couple to $S$,
  their~inclusion will not change any of the conclusions
  qualitatively.}. Since no $A$-terms are allowed by charge
conservation there is no possibility to lift the other two fields when
the first develops a VEV. Such a potential also suffers from
accidental global symmetries since there are three fields, but no
terms in the superpotential. If one hopes to overcome these problems,
the addition of extra singlets seems inevitable.

For concreteness we consider the above charge assignment with $n_{\bf
  55^\ast}=2$ and identify the singlets $S$ and $S_{\scriptscriptstyle
  L}$. Adding only one extra singlet $S_1$ will not help resolve the
axion problem. To see that, notice that since $Q_S = 1$ and
$Q_{\SD}=3/2$, there is only one term one can add to the
superpotential at the renormalizable level ($S \SD S_1$, $S S S_1$, or
$\SD \SD S_1$, but not more than one of these terms). Since there are
now three fields and one superpotential term, we are still left with
one accidental symmetry.

If we add two extra singlet fields (on top of $S$ and $\SD$), the
situation is more manageable. To remove all the accidental symmetries
we require three independent superpotential terms.  While it is
impossible to do so with cubic terms alone\footnote{The smallest
  example which contains only cubic terms requires the introduction of
  4 extra singlet fields $S_1,S_2,S_3,S_4$. The required
  superpotential contains the following 5 terms:
  $S_1S_1S_2,\,S_2S_3\SD,\,S_1S_4\SD,\,SS_3S_3,$ and $SSS_4$. The
  charges of the new fields are
  $Q_{S_1}=1/2,\,Q_{S_2}=-1,\,Q_{S_3}=-1/2,$ and $Q_{S_4}=-2$.  (The
  bilinear terms $SS_2$ and $S_1S_3$ are allowed but not needed.) As
  explained in the previous section, the reason we need so many new
  fields has to do with the fact that $Q_S$ and $Q_{\SD}$ are not
  congruent modulo 3.}, it is possible to find examples if we allow a
bilinear term. One may object to such a construction on the grounds
that it defies the original purpose of considering such models as
being free of the $\mu$ problem. However, in this case, the bilinear
term need not have any particular scale, but is there solely to give
the light scalar a large enough mass.

A simple example has two extra singlets with charge assignment,
$Q_{S_1} = -1$ and $Q_{S_2}=-1/2$. This allows for the superpotential
\begin{equation}
\label{eqn:superpotential}
W \supset \mu_{S S_1} S S_1 + y_1 S_1 S_2 \SD + y_2 S S_2 S_2 .
\end{equation}
The associated scalar potential is 
\begin{equation}
V\left(S,\SD,S_1,S_2\right) = \sum_i m_i^2 |S_i|^2 +
\frac{\gz^2}{2}\left(\sum_i Q_i|S_i|^2 \right)^2 + \sum_i |F_i|^2,
\end{equation}
where  the sum is over all four fields and the $F$-terms are given
by
\begin{eqnarray}
F_S &=& \mu_{S S_1} S_1 + y_2 S_2 S_2 \nonumber\\
F_{\SD} &=& y_1S_1 S_2 \nonumber\\
F_{S_1} &=& \mu_{S S_1} S + y_1 \SD S_2 \nonumber\\
F_{S_2} &=& y_1 \SD S_1 + 2 y_2 S S_2.
\end{eqnarray}

\subsection{Multiple Scalar Condensation}

Since the singlet fields $S$ and $\SD$ are coupled to the exotic
matter fields, it is reasonable to expect their soft masses to run
negative at low energies if the Yukawa couplings are sufficiently
large. Therefore, it is possible to destabilize the origin for at
least one of the fields. However, since both $S$ and $\SD$ have the
same sign charge, they cannot both develop a VEV in the absence of the
other fields. Suppose $\SD$ has a larger absolute value of mass
squared to charge ratio. Then, at least initially, we can set $S=0$ in
everything that follows.

As we argued above, if no flat directions are present in the limit
where only the $D$-terms are considered, then $\SD$ alone will develop
a VEV, even when $F$-terms are included. Since there are no $A$-terms
of the form $\SD \SD S_i$, no other field can develop a VEV. This is
phenomenologically unacceptable as no $\mu$ term is generated for the
MSSM and some of the exotics remain massless. To avoid that, we must
require some flat directions to be present, or in other words,
$|Q_{S_1}| m_{\SD}^2 + |Q_{\SD}| m_{S_1}^2 < 0$ and/or $|Q_{S_2}|
m_{\SD}^2 + |Q_{\SD}| m_{S_2}^2 < 0$.  For simplicity, we will mainly
consider examples in which only one is negative.

In this case, we include the $F$-terms to stabilize the potential,
considering only the resulting quartics while neglecting the bilinear
and $A$ terms. The resulting potential is
\begin{eqnarray}
V\left(S,\SD,S_1,S_2\right) &= &\sum_i m_i^2 |S_i|^2 +
\frac{\gz^2}{2}\left(\sum_i Q_i|S_i|^2 \right)^2 +\nonumber\\
&+&|y_1|^2(|S_1|^2|\SD|^2 + |\SD|^2|S_2|^2 + |S_2|^2 |S_1|^2) +
|y_2|^2 |S_2|^4 .
\end{eqnarray}
The conditions for an extremum, $\partial_{S_i} V = 0$, assuming
$\SD\ne 0$, $S_1 \ne 0$, and $S_2\ne 0$ can be written as
\begin{equation}
\left(\begin{array}{ccc}
\frac{9\gz^2}{2}& -3\gz^2 + 2 y_1^2& -\frac{3\gz^2}{2}+2y_1^2\\
-3\gz^2 + 2y_1^2&2\gz^2 &\gz^2+2y_1^2 \\
-\frac{3\gz^2}{2}+2y_1^2&\gz^2+2y_1^2 &\frac{\gz^2}{2} + 4 y_2^2
\end{array}\right)
\left(\begin{array}{c}
\SD^2\\
S_1^2\\
S_2^2
\end{array}\right)
=
-2 \left(\begin{array}{c}
m_{\SD}^2\\
m_{S_1}^2\\
m_{S_2}^2
\end{array}\right) .
\end{equation}
The solution is in general unique, but not necessarily physical, in
which case either $S_1 = 0$ or $S_2 = 0$. Instead of presenting the
solution, it is somewhat more instructive to consider the following
limits.

\medskip
\textbf{$\mathbf{y_1\ll y_2,\,\gz}$ limit}
\medskip

Assuming $|Q_{S_1}|m_{\SD}^2 + |Q_{\SD}| m_{S_1}^2 >0$, we know that
$\SD$ and $S_1$ cannot simultaneously develop a VEV and we therefore
set $S_1 =0$. Solving for $\SD$ and $S_2$ we find
\begin{equation}
|\SD|^2 = -\frac{4}{9}\frac{m_{\SD}^2}{\gz^2} - \frac{1}{18y_2^2} \left( 3m_{S_2}^2 + m_{\SD}^2\right) \quad \quad |S_2|^2 = -\frac{1}{6y_2^2} \left( 3m_{S_2}^2 + m_{\SD}^2\right) .
\end{equation}
This is the global minimum as long as $|Q_{S_2}| m_{\SD}^2 +
|Q_{\SD}| m_{S_2}^2 < 0$. As mentioned before, the dependence of the
VEV on $1/y_2^2$ is a remnant of the flat direction
$Q_{\SD}|\SD|^2 + Q_{S_2}|S_2|^2 = 0$.
This vacuum is particularly simple, and once the $A$-term for $S_2S_2
S$ is turned on, $S$ will develop a VEV and an acceptable
phenomenology results. Notice that we did not have to assume any
relation  between $y_2$ and $\gz$. As explained in the previous
section, this is a result of the fact that in this case we only have
quartic terms in the scalar potential.

\medskip
\textbf{$\mathbf{y_2\ll y_1,\,\gz}$ limit}
\medskip

In this case we have to separate the two parameter regions $y_1<\gz$ and $\gz<y_1$. If $y_1<\gz$,
$|Q_{S_1}|m_{\SD}^2 + |Q_{\SD}| m_{S_1}^2 >0$, and $|Q_{S_2}| m_{\SD}^2 +
|Q_{\SD}| m_{S_2}^2 < 0$, then we again have a minimum for $S_1=0$. Keeping the leading
terms in $1/y_1^2$,
\begin{equation}
|\SD|^2 = -\frac{1}{6y_1^2} \left( 3m_{S_2}^2 + m_{\SD}^2\right) \quad \quad |S_2|^2 = -\frac{1}{2y_1^2} \left( 3m_{S_2}^2 + m_{\SD}^2\right),
\end{equation}
similar to the previous limit. Once the $A$-term for $S_2S_2
S$ is turned on, $S$ will develop a small VEV.

If $y_1$ is much larger than $\gz$, the flat direction is so strongly
lifted that neither $S_1$ nor $S_2$ can develop a VEV. This can be
understood qualitatively as follows. After $\SD$ condenses, it can
drive the mass of $S_1$ or $S_2$ negative through the $D$ term
contribution. However, with large $F$-terms there are additional,
strictly positive contributions, which will keep the origin stable if
$y_{1} \gg \gz$.

Following similar lines as delineated above, it is possible to
construct the full phase structure of the scalar potential. In
particular, it is straightforward to find solutions for which $S_D,
S_1$, and $S_2$ are all nonzero. We will not pursue this course of
investigation and simply point out that the phases found above where
multiple fields develop a VEV are sufficient and generic.

\section{Conclusions}

In this letter we addressed two issues which are often left unchecked
in supersymmetric $\uonep$ model building: the singlet scalars'
potential must be such that no accidental global symmetries are
present, and a sufficient number of fields must develop VEVs.

The former requirement can be satisfied for $N$ fields if $N-1$ terms
can be written in the superpotential which break all the accidental
phase rotations. In some cases, this is impossible to achieve without
bilinear terms. The inclusion of such bilinears does not necessarily
reintroduce the $\mu$ problem of the MSSM since their scale is not
needed to achieve a particular vacuum structure and therefore is not
tied in with any other scale. Their purpose is simply to remove the
degeneracy and give the light boson a large enough mass.

As far as the vacuum structure is concerned, assuming that one
scalar's mass is driven negative by the soft SUSY breaking RGEs, the
following prescription emerges. If, in the absence of any $F$-terms,
no flat directions are present, then only a single field will develop
a VEV. After including $A$-terms one may destabilize the origin for
other fields if any tadpoles result from VEV insertions. If flat
directions are present, it is possible to generate a VEV for more than
one field simultaneously by stabilizing the potential with the
$F$-terms. It is then easier to ensure that all the scalars develop a
VEV once $A$-terms are included.

We illustrated these points with a particular example constructed by
Erler \cite{Erler:2000wu}.~However, we expect these considerations to
be relevant for other $\uonep$ models
whenever there are two or more SM-singlet fields with different nonzero \uonep\ charges.

{\em Acknowledgments:\/} We would like to thank Jorge de Blas Mateo,
Aviv Censor, Jens Erler, and Lian-Tao Wang for useful discussions. The
work of P.L is supported by the IBM Einstein Fellowship and by NSF
grant PHY-0503584. The work of G.P. is supported in part by the
Department of Energy grant DE-FG02-90ER40542 and by the United
States-Israel Bi-national Science Foundation grant 2002272. I.Y. is
supported by the National Science Foundation under grant PHY-0756966
and the Department of Energy under grant DE-FG02-90ER40542.

\bibliography{potential}

\providecommand{\href}[2]{#2}\begingroup\raggedright\begin{thebibliography}{10}

\bibitem{Leike:1998wr}
A.~Leike, {\it {The phenomenology of extra neutral gauge bosons}},  {\em Phys.
  Rept.} {\bf 317} (1999) 143--250,
  [\href{http://xxx.lanl.gov/abs/hep-ph/9805494}{{\tt hep-ph/9805494}}].

\bibitem{Rizzo:2006nw}
T.~G. Rizzo, {\it {$Z'$ phenomenology and the LHC}},
  \href{http://xxx.lanl.gov/abs/hep-ph/0610104}{{\tt hep-ph/0610104}}.

\bibitem{Langacker:2008yv}
P.~Langacker, {\it {The Physics of Heavy $Z'$ Gauge Bosons}},
  \href{http://xxx.lanl.gov/abs/0801.1345}{{\tt arXiv:0801.1345}}.

\bibitem{Kim:1983dt}
J.~E. Kim and H.~P. Nilles, {\it {The mu Problem and the Strong CP Problem}},
  {\em Phys. Lett.} {\bf B138} (1984) 150.

\bibitem{Suematsu:1994qm}
D.~Suematsu and Y.~Yamagishi, {\it {Radiative symmetry breaking in a
  supersymmetric model with an extra $U(1)$}},  {\em Int. J. Mod. Phys.} {\bf
  A10} (1995) 4521--4536, [\href{http://xxx.lanl.gov/abs/hep-ph/9411239}{{\tt
  hep-ph/9411239}}].

\bibitem{Cvetic:1995rj}
M.~Cvetic and P.~Langacker, {\it {Implications of Abelian Extended Gauge
  Structures From String Models}},  {\em Phys. Rev.} {\bf D54} (1996)
  3570--3579, [\href{http://xxx.lanl.gov/abs/hep-ph/9511378}{{\tt
  hep-ph/9511378}}].

\bibitem{Cvetic:1997ky}
M.~Cvetic, D.~A. Demir, J.~R. Espinosa, L.~L. Everett, and P.~Langacker, {\it
  {Electroweak breaking and the mu problem in supergravity models with an
  additional $U(1)$}},  {\em Phys. Rev.} {\bf D56} (1997) 2861--2885,
  [\href{http://xxx.lanl.gov/abs/hep-ph/9703317}{{\tt hep-ph/9703317}}].

\bibitem{Accomando:2006ga}
E.~Accomando {\em et.~al.}, {\it {Workshop on CP studies and non-standard Higgs
  physics}},  \href{http://xxx.lanl.gov/abs/hep-ph/0608079}{{\tt
  hep-ph/0608079}}.

\bibitem{Langacker:2007ac}
P.~Langacker, G.~Paz, L.-T. Wang, and I.~Yavin, {\it {$Z'$-mediated
  Supersymmetry Breaking}},  {\em Phys. Rev. Lett.} {\bf 100} (2008) 041802,
  [\href{http://xxx.lanl.gov/abs/0710.1632}{{\tt arXiv:0710.1632}}].

\bibitem{Langacker:2008ip}
P.~Langacker, G.~Paz, L.-T. Wang, and I.~Yavin, {\it {Aspects of Z'-mediated
  Supersymmetry Breaking}},  {\em Phys. Rev.} {\bf D77} (2008) 085033,
  [\href{http://xxx.lanl.gov/abs/0801.3693}{{\tt arXiv:0801.3693}}].

\bibitem{Hewett:1988xc}
J.~L. Hewett and T.~G. Rizzo, {\it {Low-Energy Phenomenology of Superstring
  Inspired $E(6)$ Models}},  {\em Phys. Rept.} {\bf 183} (1989) 193.

\bibitem{Langacker:1998tc}
P.~Langacker and J.~Wang, {\it {$U(1)' $ symmetry breaking in supersymmetric
  $E(6)$ models}},  {\em Phys. Rev.} {\bf D58} (1998) 115010,
  [\href{http://xxx.lanl.gov/abs/hep-ph/9804428}{{\tt hep-ph/9804428}}].

\bibitem{Erler:2000wu}
J.~Erler, {\it {Chiral models of weak scale supersymmetry}},  {\em Nucl. Phys.}
  {\bf B586} (2000) 73--91, [\href{http://xxx.lanl.gov/abs/hep-ph/0006051}{{\tt
  hep-ph/0006051}}].

\bibitem{Morrissey:2005uz}
D.~E. Morrissey and J.~D. Wells, {\it {The tension between gauge coupling
  unification, the Higgs boson mass, and a gauge-breaking origin of the
  supersymmetric mu-term}},  {\em Phys. Rev.} {\bf D74} (2006) 015008,
  [\href{http://xxx.lanl.gov/abs/hep-ph/0512019}{{\tt hep-ph/0512019}}].

\bibitem{Kang:2007ib}
J.~Kang, P.~Langacker, and B.~D. Nelson, {\it {Theory and Phenomenology of
  Exotic Isosinglet Quarks and Squarks}},  {\em Phys. Rev.} {\bf D77} (2008)
  035003, [\href{http://xxx.lanl.gov/abs/0708.2701}{{\tt arXiv:0708.2701}}].

\bibitem{Georgi:1986df}
H.~Georgi, D.~B. Kaplan, and L.~Randall, {\it {Manifesting The Invisible Axion
  At Low-Energies}},  {\em Phys. Lett.} {\bf B169} (1986) 73.

\bibitem{Long:2003ta}
J.~C. Long and J.~C. Price, {\it {Current short-range tests of the
  gravitational inverse square law}},  {\em Comptes Rendus Physique} {\bf 4}
  (2003) 337--346, [\href{http://xxx.lanl.gov/abs/hep-ph/0303057}{{\tt
  hep-ph/0303057}}].

\bibitem{Adelberger:2003zx}
E.~G. Adelberger, B.~R. Heckel, and A.~E. Nelson, {\it {Tests of the
  gravitational inverse-square law}},  {\em Ann. Rev. Nucl. Part. Sci.} {\bf
  53} (2003) 77--121, [\href{http://xxx.lanl.gov/abs/hep-ph/0307284}{{\tt
  hep-ph/0307284}}].

\bibitem{Amsler:2008zz}
{\bf Particle Data Group} Collaboration, C.~Amsler {\em et.~al.}, {\it {Review
  of particle physics}},  {\em Phys. Lett.} {\bf B667} (2008) 1.

\bibitem{Giudice:1988yz}
G.~F. Giudice and A.~Masiero, {\it {A Natural Solution to the mu Problem in
  Supergravity Theories}},  {\em Phys. Lett.} {\bf B206} (1988) 480--484.

\bibitem{Chung:2003fi}
D.~J.~H. Chung {\em et.~al.}, {\it {The soft supersymmetry-breaking Lagrangian:
  Theory and applications}},  {\em Phys. Rept.} {\bf 407} (2005) 1--203,
  [\href{http://xxx.lanl.gov/abs/hep-ph/0312378}{{\tt hep-ph/0312378}}].

\bibitem{Erler:2002pr}
J.~Erler, P.~Langacker, and T.-j. Li, {\it {The $Z - Z'$ mass hierarchy in a
  supersymmetric model with a secluded $U(1)'$-breaking sector}},  {\em Phys.
  Rev.} {\bf D66} (2002) 015002,
  [\href{http://xxx.lanl.gov/abs/hep-ph/0205001}{{\tt hep-ph/0205001}}].

\bibitem{Cleaver:1997nj}
G.~Cleaver, M.~Cvetic, J.~R. Espinosa, L.~L. Everett, and P.~Langacker, {\it
  {Intermediate scales, mu parameter, and fermion masses from string models}},
  {\em Phys. Rev.} {\bf D57} (1998) 2701--2715,
  [\href{http://xxx.lanl.gov/abs/hep-ph/9705391}{{\tt hep-ph/9705391}}].

\end{thebibliography}\endgroup
\bibliographystyle{JHEP}
\end{document}